\documentstyle[12pt]{article}
\begin{document}
\renewcommand{\theequation}{\thesection.\arabic{equation}}
\vskip 2cm
\title{String Spreading on  Black Hole Horizon}
\author{\\
  A.L. Larsen${}^{|} {}^{1}$ and A. Nicolaidis${}^{||} {}^{2}$}
\maketitle
\noindent
$^{1}${\em Physics Department, University of Odense, Campusvej 55, 5230
Odense M,
Denmark}\\
$^{2}${\em Theoretical  Physics Department, University of Thessaloniki,
54006 Thessaloniki, Greece}
\vskip 6cm
\noindent
$^{|}$Electronic address: all@fysik.ou.dk\\
$^{||}$Electronic address: nicolaid@ccf.auth.gr
\newpage
\begin{abstract}
\baselineskip=1.5em
\hspace*{-6mm}The phenomenon of string spreading on the black hole
horizon, as originally discussed by Susskind, is considered in the {\it exact}
curved Schwarzschild background. We consider
an  oscillating string encircling the black hole and contracting
towards the horizon.
We then compute the angular and radial spreading of the string, as seen by
a static observer at
spatial infinity using fixed finite resolution time.
   Within our case study we find that there is indeed a  spreading of the
string in the angular
direction, such that the string eventually covers the whole horizon. However,
regarding the radial direction, we find that
Lorentz-contraction suppresses the radial string spreading.
\end{abstract}
\noindent
\newpage
\section{Introduction}
\setcounter{equation}{0}
Some of the most important problems in theoretical high energy physics
today, are related
to the quantum properties of matter near a black hole horizon (not to speak
about the
quantum properties of the black holes themselves). Especially the problem
of the microscopic
explanation of the black hole entropy and the problem of information loss,
have attracted
a lot of interest in recent years.

In many approaches to these problems, the so-called
"stretched horizon" plays an important role. For an outside observer
the stretched horizon appears as  a membrane placed near the event horizon
of the
black hole, endowed with mechanical, electrical and thermal properties
(see for instance \cite{TPM}, and references given therein).
The interactions of the stretched horizon
with the outside world can be thought of as arising from the boundary
conditions that must be implemented. 't Hooft \cite{tH} and Susskind
\cite{LS1} further suggested that the stretched horizon exists also as
a collection of quantum mechanical microscopic degrees of freedom which
can absorb, store and re-emit any quantum mechanical information which
falls into the black hole. The above description by an external observer
is complementary \cite{tH, LS1, LTU} to the description provided by a
freely falling
observer, who crosses the horizon and ends up at the singularity.
Within this spirit, Susskind \cite{susk1} studied the spreading of a string
approaching the event horizon of a black hole, as seen by a distant observer.
Later is was argued by Mezhlumian et al \cite{mezh}, that string
thermalization takes place, and
eventually the information carried by the string is re-emitted as
thermal Hawking
radiation. To obtain
these results, quantitative results from Rindler or Minkowski
spacetime \cite{susk2}
were more or less directly taken over to the curved spacetime region near a
black hole. For
instance, in \cite{mezh}, the spherical horizon region is approximated by a
plane, corresponding to the case of an infinite mass black hole.

The problem of the classical and quantum propagation of strings in a black
hole background
is rather complicated and generally not solvable. This is due to the highly
non-linear nature of the equations of motion and the subsequent absence of a
light-cone formulation (when using conformal gauge). Therefore, in the
above mentioned papers a
number of simplifying assumptions were made. Namely, it was assumed that
the infalling string
is very small in both the radial and angular directions, compared to the
curvature radius of
the black hole background. And then the horizon region of the black hole
spacetime was
approximated first by "radial" Rindler space (with
$R^2\times S^{D-2}$ geometry) and eventually by "Cartesian" Rindler space
(with flat
$R^{D}$ geometry). This allowed a light-cone formulation, and the equations
of motion could
be solved. However our ultimate goal, to study an extended string covering
the whole horizon, renders the above assumptions and approximations
inconsistent and inappropriate.

The purpose of the present paper is  to address again this important issue,
avoiding unwarranted approximations and carrying
out all computations directly in the {\it exact} curved spacetime
metric of a black hole.
Everything will  be expressed using Schwarzschild
coordinates
corresponding to a static asymptotic observer, in order to clarify what the
static
asymptotic observer
actually sees.
As already mentioned, there is no hope that we can solve the general string
equations of motion
in the exact curved Schwarzschild background. However, it is
possible to find a special
solution, and then more general solutions can be obtained by linearization
of the equations of
motion around this special solution. In this way we have complete
control of the approximation involved.

Our physical setup will however be somewhat different from that discussed
in references
\cite{susk1,mezh}. In the original papers \cite{susk1,mezh}, a small string
was considered. The string was
initially far outside the black hole, but then approached the horizon while
oscillating
around its center of mass.
We will consider instead a macroscopic closed string initially encircling
the black hole horizon. The string is oscillating around some average shape,
which is a
circle, but its overall average dynamics is a contraction towards the
horizon. The string
center of mass is therefore not part of the string, but is at all times
located at the
singularity of the black hole.
For such string configurations it is possible to compute the
string spreading in the radial and angular directions, while consistently
working in the exact
curved Schwarzschild background.

In the next section, we present the
circular string solution in  the
Schwarzschild background and the perturbations around it.
In section 3, we
examine
the string spreading, as the black hole horizon is approached, in both the
angular and the radial directions. Finally in section 4, we present our
conclusions.

\section{Circular String Oscillations}
\setcounter{equation}{0}
Our starting point is the Polyakov action
\begin{equation}
S_{(0)}=\frac{-1}{4\pi\alpha'}\int d\tau d\sigma\;\sqrt{-h}\;h^{AB}G_{AB},
\end{equation}
where $h_{AB}$ is the Polyakov metric while $G_{AB}$ is the induced metric
on the string
world-sheet
\begin{equation}
G_{AB}=g_{\mu\nu}x^\mu_{,A}x^\nu_{,B}.
\end{equation}
Here $(A, B)=(0, 1)$ are  the world-sheet indices while $(\mu, \nu)=(0, 1,
2, 3)$ are the
spacetime indices. For simplicity we consider strings in a 4-dimensional
spacetime, but all
our results are easily generalized to arbitrary dimensions.

The  equations of motion corresponding to the action (2.1),
\begin{equation}
G_{AB}=\frac{1}{2}h_{AB}G^C\;_C,
\end{equation}
\begin{equation}
\Box x^\mu+h^{AB}\Gamma^\mu_{\rho\sigma}x^\rho_{,A}x^\sigma_{,B}=0,
\end{equation}
contrary to the case of flat Minkowski space, cannot generally be solved in
a curved spacetime
with metric
$g_{\mu\nu}$. In most cases it is however possible to find exact special
solutions
$(h_{AB},\;x^\mu)$. In particular, in the Schwarzschild black hole
spacetime a number of exact
special solutions are known \cite{fro, all, sanchez, vega, hendy}. More
general solutions
can then be found by considering  linearized perturbations $(\delta
h_{AB},\; \delta x^\mu)$
around such exact special solutions. Moreover, since we are interested only
in physical
(transverse) perturbations, $\delta x^\mu$ can be expanded on a set of
unit-vectors normal to
the world-sheet of the exact special solution
\begin{equation}
\delta x^\mu=n^\mu_{i}\Phi^{i},
\end{equation}
where
\begin{equation}
g_{\mu\nu}n^\mu_{i}n^\nu_{j}=\delta_{ij},\;\;\;\;\;\;\;\;g_{\mu\nu}n^\mu_{i}
x^\nu_{,A}=0.
\end{equation}
It is then straightforward to show that the fields $\Phi^{i}$ ($i=1, 2$)
are governed by the
action
\cite{guven, carter, frolov}
\begin{equation}
S_{(2)}=\frac{1}{2\pi\alpha'}\int d\tau d\sigma\;\sqrt{-h}\;\Phi^{i}
\left[h^{AB}{\cal D}_{ikA} {\cal D}_{kjB}-{\cal V}_{ij}\right]\Phi^{j},
\end{equation}
where
\begin{equation}
{\cal D}_{ikA}=\delta_{ik}\nabla_A+\mu_{ikA},
\end{equation}
\begin{equation}
{\cal V}_{ij}=h^{AB}R_{\mu\rho\sigma\nu}x^\mu_{,A}x^\nu_{,B} n^\rho_{i}
n^\sigma_{j}-
\frac{2}{G^C\;_C}\Omega_{iAB}\Omega_j\;^{AB}.
\end{equation}
Here we also introduced the extrinsic curvature and torsion of the
world-sheet
\begin{equation}
\Omega_{iAB}=g_{\mu\nu}n^\mu_{i} x^\rho_{,A}\nabla_\rho x^\nu_{,B}\;\;\;\;
, \;\;\;\;
\mu_{ijA}=g_{\mu\nu}n^\mu_{i} x^\rho_{,A}\nabla_\rho n^\nu_{j},
\end{equation}
as well as the Riemann tensor $R_{\mu\rho\sigma\nu}$ of the curved spacetime.

In the present paper we take as the exact special (unperturbed) solution a
circular string.
That is, we take $h_{AB}=\eta_{AB}$ (gauge choice) and make the ansatz
\begin{equation}
x^{0}=t=t(\tau),\;\;\;\;x^{1}=r=r(\tau),\;\;\;\;
x^{2}=\theta=\pi/2,\;\;\;\;x^{3}=\phi=\sigma,
\end{equation}
describing a circular string with time-dependent radius in the equatorial
plane of the
Schwarzschild black hole
\begin{equation}
ds^2=-(1-2M/r)dt^2+(1-2M/r)^{-1}dr^2+r^2(d\theta^2+\sin^2\theta d\phi^2).
\end{equation}
The  equations of motion become
\begin{equation}
\dot{t}=\frac{E}{1-2M/r},
\end{equation}
\begin{equation}
\dot{r}^2=E^2-r^2(1-2M/r),
\end{equation}
and they are solved by \cite{all}
\begin{equation}
t(\tau)=E\tau+2M\log\left|\frac{\tan{(\tau/2)}+\delta}{\tan{(\tau/2)}-\delta}
\right|,
\end{equation}
\begin{equation}
r(\tau)=M+\sqrt{M^2+E^2}\;\cos(\tau),
\end{equation}
where $E$ is an integration constant and $\delta=(\sqrt{M^2+E^2}-M)/E$.
Physically this
solution describes a macroscopic circular string, encircling the black hole
in the equatorial
plane, starting at
$\tau=0$ with maximal radius
$r_{max}$
\begin{equation}
r_{max}=r(\tau=0)=M+\sqrt{M^2+E^2}.
\end{equation}
It then contracts and crosses the horizon at finite world-sheet time $\tau_h$
\begin{equation}
\tau_h=\tau(r=2M)=\arccos(M/\sqrt{M^2+E^2}\;)\in\;]0,\pi/2[\;,
\end{equation}
which of course corresponds to infinite coordinate time $t(\tau_h)=\infty$.
The string
eventually  falls into the singularity at world-sheet time $\tau_0$
\begin{equation}
\tau_0=\tau(r=0)=\arccos(-M/\sqrt{M^2+E^2}\;)\in\;]\pi/2,\pi[\;.
\end{equation}
The integration constant $E$ has the physical
interpretation of the constant
energy of the string (in suitable units).

It is important to notice that the string at the horizon behaves as a
null string \cite{HdV} in the radial direction. For a null
string the string tension is set to zero and the string appears as a
collection of individual massless particles, each following its own null
geodesic line. We suggest that this is a general feature: any kind of matter
at the horizon looks like a collection of massless particles. This
{\it horizon universality} is reminiscent of the universality encountered
in the parton model (for early work on the parton model see \cite{KS}).
In a hard collision involving
a hadron (or hadrons) we probe the short distance structure of a hadron, and
the partonic degrees of freedom, universal for all collisions, are revealed.
Non-leading corrections to the parton model are sensitive to the specific
character of the collision involved. In our case the black hole geometry
provides the high energy required to probe the short distance behavior of
the infalling matter, and it appears that any kind of matter at the
event horizon of a black hole, that is in the ultraviolet limit, looks
identical. The horizon universality is further reflected in the emitted
thermal Hawking radiation \cite{Hawking,unruh}, which involves particles
very close to the horizon and ignores the initial state of the infalling
matter. We suspect that if we take into account the whole evolution of the
infalling matter before reaching the horizon, then the emitted radiation
will be non-thermal and will contain all the relevant information of the
initial state (in close analogy to non-leading corrections to the parton
model). The above emerging picture is in accordance with the line of
reasoning advocated by 't Hooft \cite{tH}. Our concrete example
also realizes the holographic principle \cite{tH, Sholo}: null
rays connect the event horizon with the outside world, as well as
with the interior of the black hole.

We now consider perturbations around the solution (2.15)-(2.16).
The two physical polarizations of oscillations (2.5) are radial and angular
oscillations,
respectively, in the directions of the unit normal vectors
\begin{equation}
n^\mu_{r}=\left(
\frac{\dot{r}}{r-2M},\;\frac{\dot{t}}{r^2}(r-2M),\;0,\;0\right),
\end{equation}
\begin{equation}
n^\mu_{\theta}=\left( 0,\;0,\;\frac{1}{r},\;0\right).
\end{equation}
After a little algebra, the action (2.7) becomes
\begin{equation}
S_{(2)}=\frac{1}{2\pi\alpha'}\int d\tau d\sigma \; \left\{
\Phi_{r}\left(
-\partial^2_\tau+\partial^2_\sigma+\frac{M}{r}+\frac{2E^2}{r^2}\right)\Phi_{r}
+
\Phi_{\theta}\left(
-\partial^2_\tau+\partial^2_\sigma+\frac{M}{r}\right)\Phi_{\theta} \right\} .
\end{equation}
When the string approaches the horizon, the action  reduces to
\begin{equation}
S_{(2)}=\frac{1}{2\pi\alpha'}\int d\tau d\sigma\; \left\{
\dot{\Phi}^2_{r}-\Phi'^2_{r}
+\dot{\Phi}^2_{\theta}-\Phi'^2_{\theta}+\frac{1}{2}\left(1+\frac{E^2}{M^2}
\right)\Phi^2_{r}+
\frac{1}{2}\Phi^2_{\theta}\right\}.
\end{equation}
The equations of motion corresponding to the action (2.23) are solved by
\begin{equation}
\Phi_{r}=\frac{\sqrt{\alpha'}}{2}\sum_{n>0}\frac{1}{\sqrt{n\Omega_n}}\left[
a_n e^{-in(\Omega_n\tau-\sigma)}+\tilde{a}_n e^{-in(\Omega_n\tau+\sigma)} +
c.c.\right],
\end{equation}
\begin{equation}
\Phi_{\theta}=\frac{\sqrt{\alpha'}}{2}\sum_{n>0}\frac{1}{\sqrt{n\omega_n}}\left[
b_n e^{-in(\omega_n\tau-\sigma)}+\tilde{b}_n e^{-in(\omega_n\tau+\sigma)} +
c.c.\right],
\end{equation}
where $(\Omega_n,\;\omega_n)$ are given by
\begin{equation}
\Omega_n=\sqrt{1-\frac{1+E^2/M^2}{2n^2}}\;\;\;\;,\;\;\;\;\omega_n=\sqrt{1-\frac{
1}{2n^2}}.
\end{equation}
Equations (2.24)-(2.25) together with eqs.(2.20)-(2.21) and eq.(2.5) thus
provide the first
order perturbations around the circular string (2.15)-(2.16), when the
string is
in the vicinity of
the black hole horizon. And it should be stressed again that we are
treating the Schwarzschild
background {\it exactly}; no Rindler approximations are involved.

From
the action (2.23) also follows that the momenta conjugate to
$(\Phi_{r},\;\Phi_{\theta})$ are given by
\begin{equation}
\Pi_{r}=\frac{\delta
S_{(2)}}{\delta\dot{\Phi}_{r}}=\frac{1}{\pi\alpha'}\dot{\Phi}_{r}
\;\;\;\;,\;\;\;\;
\Pi_{\theta}=\frac{\delta
S_{(2)}}{\delta\dot{\Phi}_{\theta}}=\frac{1}{\pi\alpha'}\dot{\Phi}_{\theta}
\end{equation}
with equal-time Poisson-brackets
\begin{equation}
\{ \Pi_{r},\;\Phi_{r} \}=\{ \Pi_{\theta},\;\Phi_{\theta}
\}=-\delta(\sigma-\sigma').
\end{equation}
It follows that $(a_n,\tilde{a}_n,b_n,\tilde{b}_n)$ are properly normalized
"oscillators"
\begin{equation}
\{a_n,a^*_m \}=\{\tilde{a}_n,\tilde{a}^*_m \}=\{b_n,b^*_m
\}=\{\tilde{b}_n,\tilde{b}^*_m
\}=-i\delta_{nm},
\end{equation}
that is, at the quantum level, the oscillators
$(a_n,\tilde{a}_n,b_n,\tilde{b}_n)$ will become
the standard creation and annihilation operators.

Notice also that the zero-modes were
eliminated from the summations in eqs.(2.24)-(2.25). As discussed in reference
\cite{vilenkin}, both the zero-modes as well as the $n=1$-modes should in
fact be eliminated
since they do not represent "true" oscillations of a circular string; they
merely describe
rigid translations and rotations that do not change the shape of the
string. Elimination of
the zero-modes, in particular, ensures that the frequencies (2.26) are real
for sufficiently
small values of the integration constant $E/M$.
\section{String Spreading}
\setcounter{equation}{0}
In this section we shall consider the spreading of the circular string, as
seen by a static
asymptotic observer. Up to first order perturbations around the circular
string, we have
\begin{equation}
r(\tau,\sigma)=r(\tau)+\frac{E}{r(\tau)}\Phi_{r}(\tau,\sigma),
\end{equation}
\begin{equation}
\theta(\tau,\sigma)=\frac{\pi}{2}+\frac{1}{r(\tau)}\Phi_{\theta}(\tau,\sigma),
\end{equation}
where $r(\tau)$ is the unperturbed string (2.16).
Then the radial and angular spreading due to the oscillations are
\begin{equation}
R_{r}\equiv r(\tau,\sigma)-r(\tau)=\frac{E}{r(\tau)}\Phi_{r}(\tau,\sigma),
\end{equation}
\begin{equation}
R_\theta\equiv
r(\tau,\sigma)\cos[\theta(\tau,\sigma)]=\Phi_{\theta}(\tau,\sigma).
\end{equation}
Using eqs.(2.24)-(2.25), the (square of the) average spreading of a
circular string near the
horizon is then given by
\begin{equation}
<R^2_{r}>=\frac{1}{2\pi}\int_{0}^{2\pi}
d\sigma\;\frac{E^2}{4M^2}[\Phi_{r}(\tau,\sigma)]^2,
\end{equation}
\begin{equation}
<R^2_{\theta}>=\frac{1}{2\pi}\int_{0}^{2\pi}
d\sigma\;[\Phi_{\theta}(\tau,\sigma)]^2,
\end{equation}
which leads to
\begin{equation}
<R^2_{r}>=\frac{E^2\alpha'}{16M^2}\sum_{n>0}\frac{1}{n\Omega_n}\left[ a_n a_n^*
+\tilde{a}_n \tilde{a}_n^* +2a_n \tilde{a}_n e^{-2in\Omega_n
\tau}+c.c.\right] ,
\end{equation}
\begin{equation}
<R^2_{\theta}>=\frac{\alpha'}{4}\sum_{n>0}\frac{1}{n\omega_n}
\left[ b_n b_n^* +\tilde{b}_n \tilde{b}_n^* +2b_n \tilde{b}_n
e^{-2in\omega_n \tau}+c.c.\right] .
\end{equation}
The constants $(a_n, b_n, \tilde{a}_n, \tilde{b}_n)$ are of the order $1$,
so that the
summations in eqs.(3.7)-(3.8) formally diverge logaritmically. But we shall
now argue that
the infinite summations must be truncated for physical reasons.

Assume that the asymptotic
observer  watches the contracting oscillating string with a fixed finite
resolution time
$\epsilon$ measured in coordinate time
$t$. Then he will be able to see frequencies $\nu$ fulfilling the inequality
\begin{equation}
\nu<\nu_{t}\equiv \frac{1}{\epsilon}
\end{equation}
These frequencies correspond to frequencies $\nu_\tau$ in world-sheet time
$\tau$
\begin{equation}
\nu_{t}\frac{dt}{d\tau}=\nu_{\tau}\;\;\;\Leftrightarrow\;\;\;\;\nu_\tau=E(1-2M/r
)^{-1}\;\nu_{t},
\end{equation}
That is to say, as the string approaches the horizon, the asymptotic
observer will
see more and more oscillation modes, even though his resolution time is
fixed and finite. This
is of course nothing but the standard gravitational redshift effect.

However, we must also take into account the Lorentz-contraction of the
radial oscillations.
The Lorentz-contraction factor is
\begin{equation}
\gamma^{-1}=\sqrt{1-v_p^2}\;,
\end{equation}
where $v_p$ is the physical speed of the contraction of the circular string
\begin{equation}
v_p=\frac{dl_{proper}}{d\tau_{proper}}=\left(
1-\frac{2M}{r}\right)^{-1}\frac{dr}{dt}=
\left(
1-\frac{2M}{r}\right)^{-1}\frac{dr}{d\tau}\frac{d\tau}{dt}=\frac{1}{E}\sqrt{E^2-
r(r-2M)}.
\end{equation}
Such that
\begin{equation}
\gamma^{-1}=\frac{r}{E}\sqrt{1-2M/r}.
\end{equation}
Thus altogether the asymptotic observer ($AO$) will see the following  when
the string
approaches the horizon
\begin{equation}
<R^2_{r}>_{AO}=\gamma^{-2}\frac{E^2\alpha'}{16M^2}\sum^{\nu_\tau}\frac{1}
{n\Omega_n}\left[a_n
a_n^* +\tilde{a}_n\tilde{a}_n^*+2a_n\tilde{a}_n
e^{-2in\Omega_n\tau}+c.c.\right],
\end{equation}
\begin{equation}
<R^2_{\theta}>_{AO}=\frac{\alpha'}{4}\sum^{\nu_\tau}\frac{1}{n\omega_n}\left[b_n
b_n^*
+\tilde{b}_n\tilde{b}_n^*+2b_n\tilde{b}_n e^{-2in\omega_n\tau}+c.c.\right].
\end{equation}
That is to say,
\begin{equation}
|R_{r}|_{AO}\sim \left[ \frac{\alpha'}{4}\left( 1-\frac{2M}{r}\right)
\log\left( \frac{E}{\epsilon}(1-2M/r)^{-1}\right) \right] ^{1/2}
\;\rightarrow\;\;0\;\;,\;\;\;\;r\rightarrow 2M
\end{equation}
\begin{equation}
|R_{\theta}|_{AO}\sim \left[ \frac{\alpha'}{4}
\log\left( \frac{E}{\epsilon}(1-2M/r)^{-1}\right) \right] ^{1/2}
\;\rightarrow\;\;\sqrt{\frac{\alpha'}{4}\left|
\log(1-2M/r)\right|}\;\;,\;\;\;\;r\rightarrow 2M
\end{equation}
We therefore find that there is a string spreading in the angular
direction, while Lorentz-contraction kills the corresponding radial
spreading. Our asymptotic observer will see that the string, as it is
approaching the distance $r=2M$ (to be reached at infinite coordinate time),
wraps around the (2-dimensional) horizon of the  black hole. Thus, for an
external observer,
the information carried by the string is absorbed by the entire area
of the event horizon.

\section{Conclusion}
\setcounter{equation}{0}
One of the most intriguing subjects in physics is the behavior of matter
nearby a black hole (usually exemplified by the problem of black hole
entropy and the problem of information loss). It is not clear if the
existing theoretical framework suffices to address the issue, or if a major
revision of current concepts is necessary. One promising approach evolves
around ideas proposed and elaborated by 't Hooft \cite{tH} and Susskind
\cite{LS1}. According to these ideas, all the information of infalling
matter is stored in the stretched horizon and is further re-emitted via
a Hawking-type radiation. The external viewpoint and the comoving
viewpoint are complementary to each other \cite{tH,LS1, LTU}. Along these
lines Susskind \cite{susk1}  provided an  analysis of the external viewpoint
within
string theory: how an infalling string into a black hole appears to an
asymptotic observer. He concluded that the stringy degrees of freedom
spread and cover the area of the event horizon.

In the present paper we studied the same problem in a different context,
treating exactly the curved Schwarzschild background.
We have considered a macroscopic
oscillating string, encircling a Schwarzschild black hole, and its
subsequent contraction
towards the horizon. For such string configurations we  were able  to
obtain analytic expressions for the
angular and radial
spreading, as seen by a static observer at spatial infinity. We took into
account both the finite resolution time of the observer and the
Lorentz-contraction factor.

We found a logarithmic growth of both the angular and radial string spreading.
However the radial spreading is suppressed by the stronger Lorentz-contraction.
 On the other hand Susskind \cite{susk1} and
Mezhlumian et al. \cite{mezh} obtained a  radial spreading growing
quadratically
and persisting even
after including the Lorentz-contraction effect. This difference might be
due to the different
physical configurations under study: for instance, the
radial
oscillations are in some sense "longitudinal" in the original situation
\cite{susk1,mezh}, while in our case they are actually "transverse".
The difference might be related also to the simplifications
employed in \cite{susk1,mezh} concerning the Rindler approximations.
In both studies, however, the general
picture emerges  of a string
covering eventually the whole black hole horizon.

Furthermore, our concrete physical example allowed us to establish the
universal behavior of the infalling matter at the event horizon and lend
further support to existing theoretical approaches to the information
loss problem.

\vskip 1.5 cm
\hspace*{-6mm}{\bf Acknowledgements}:\\
This work was supported by the Greek-Danish exchange program (Greek
Ministry of Education - Danish Rektorkollegiet). One of us (AN) would like
to thank the Fysisk Institut of Odense University for the warm hospitality
extended to him, during his stay in Odense.

\end{document}